\documentclass[conference]{IEEEtran}
\IEEEoverridecommandlockouts
\usepackage{indentfirst}
\usepackage{amsfonts,amsmath,amssymb}
\usepackage{epsfig}
\usepackage{graphicx}
\usepackage{epstopdf}
\usepackage{times}
\usepackage{relsize}
\usepackage{mathtools}
\usepackage{cases}
\usepackage{pifont}

\usepackage{bm}
\usepackage{xcolor}
\usepackage{comment}
\usepackage{cite}

\newtheorem{Theorem}{Theorem}

\newtheorem{Remark}{Remark}
\newtheorem{Lemma}{Lemma}

\begin{document}
	
	\title{Optimal Index Assignment for Scalar Quantizers and M-PSK via a Discrete Convolution-Rearrangement Inequality
		\thanks{The work was supported by the Hong Kong Research Grants Council under project no. GRF 16233816.}
	}
	\author{%
		\IEEEauthorblockN{Yunxiang Yao and Wai Ho Mow}
		\IEEEauthorblockA{Department of Electronic and Computer Engineering\\The Hong Kong University of Science and Technology\\
			Email: yyaoaj@ust.hk, eewhmow@ust.hk}
	}
	\maketitle
	
	\begin{abstract}
		This paper investigates the problem of finding an optimal nonbinary index assignment from \(M\) quantization levels of a maximum entropy scalar quantizer to \(M\)-PSK symbols transmitted over a symmetric memoryless channel with additive noise following decreasing probability density function (such as the AWGN channel) so as to minimize the channel mean-squared distortion. The so-called zigzag mapping under maximum-likelihood (ML) decoding was known to be asymptotically optimal, but the problem of determining the optimal index assignment for any given signal-to-noise ratio (SNR) is still open. Based on a generalized version of the Hardy-Littlewood convolution-rearrangement inequality, we prove that the zigzag mapping under ML decoding is optimal for all SNRs. It is further proved that the same optimality results also hold under minimum mean-square-error (MMSE) decoding. Numerical results are presented to verify our optimality results and to demonstrate the performance gain of the optimal \(M\)-ary index assignment over the state-of-the-art binary counterpart for the case of \(8\)-PSK over the AWGN channel. 
	\end{abstract}

	\section{Introduction}

	Index assignment (IA) is a low-complexity approach for joint source-channel coding design. The classical binary IA problem aims to assign bit labels to the quantization code vectors in a way to ensure that any two code vectors with a small Euclidean distance have their corresponding bit labels close in the Hamming space. Therefore, if the transmitted bit labels are corrupted by noise and decoded erroneously at the receiver, the resulting distortion may not be too large. Some important results on the optimal binary IA problem are well-known. For the maximum entropy scalar quantizers and the binary symmetric channel (BSC), the \emph{natural binary code} (NBC) is an optimal IA for all crossover probabilities, in the sense of minimizing the channel mean-squared distortion (MSD) \cite{mclaughlin1995optimal,farber2004quantizers}. For general quantizers and discrete memoryless channels (DMC) with nonbinary channel symbols, lower bounds for the channel MSD were studied in \cite{boundIT05,wu2010bound}. Generally speaking, finding the optimal IA is known to be NP-hard \cite{boundIT05}. To the best of our knowledge, there are few previous works on the optimal IA problem, except for the aforementioned binary case and the specific nonbinary case to be described next.
	
	In \cite{chan2005index}, the nonbinary IA problem of mapping the $M$ levels of a maximum entropy scalar quantizer to the $M$-ary phase shift keying ($M$-PSK) symbols transmitted over the additive white Gaussian noise (AWGN) channel was addressed. It was proved that the so-called \emph{zigzag mapping} constructed therein is an asymptotically optimal IA at a sufficiently high signal-to-noise ratio (SNR) when a maximum-likelihood (ML) decoder is used. \textcolor{black}{As an extension of the work, in \cite{nmc2020} the authors proposed a near-optimal index assignment scheme for $M^L$-level quantizers to $M$-PSK symbols at a sufficiently high SNR.} \textcolor{black}{Besides the setting under ML decoding, the performance of the zigzag mapping for $M$-PSK with minimum mean-square-error (MMSE) decoding at a sufficiently high SNR was investigated in \cite{qiao2010scalar}.} However, the corresponding optimal IA problem for any given SNR is still open. In particular, it is unclear if different IA constructions are needed for different SNRs.
	
	\textcolor{black}{Rearrangement inequalities indicate permutations of two or more functions or sets that optimize an objective function involving them.} They are powerful tools in function analysis and are widely used in the proof of other inequalities \cite{wang2014beyond}. For instance, classical rearrangement inequalities have been applied to prove the existence and uniqueness of the ground states of the Schr{\"o}dinger equation in quantum mechanics \cite{choquard2007one}. For a convolution involving three continuous functions, the \emph{Riesz convolution-rearrangement inequality} characterizes the rearrangements of the three functions that maximize the convolution \cite{riesz1930}. It has found applications in information theory and communication problems, such as the network optimization \cite{hajek2008paging}, power entropy inequality \cite{wang2014beyond}, and Cover's problem for Gaussian relay channels \cite{Cover2019}. Its discrete version was first developed by Hardy and Littlewood \cite[Theorem 371]{hardy1952inequalities} for characterizing the rearrangement of two sets. \textcolor{black}{The inequality was extended to prime cyclic groups in \cite{lev2001linear} and then it was applied to the proof of discrete entropy inequalities in \cite{madiman2017entropy}.} In \cite{pruss1998discrete}, the Hardy-Littlewood convolution-rearrangement inequality was also generalized on discrete metric spaces. Recently, another discrete version of the Riesz rearrangement inequality on Hamming sphere was derived and applied to solve Cover's problem for the binary symmetric relay channel \cite{barnes2017solution}.  Furthermore, many results in majorization theory which plays an important role in optimization are established by using rearrangement inequalities \cite{ando1989majorization}. Therefore, it is very interesting to investigate rearrangement inequalities with applications to coding and information theory. In this work, we relate the IA problem for $M$-PSK under both ML decoding and MMSE decoding to a rearrangement inequality to settle the aforementioned problem for any given SNR.
	
	\textcolor{black}{
		The paper is structured as follows. In the next section, we state the problem formulation of index assignment for $M$-PSK under ML decoding. In Section III, we provide a discrete convolution-rearrangement inequality and apply it to the IA problem. In Section IV, we show that the optimal index assignment for $M$-PSK under MMSE decoding can also be proved by the inequality. Finally, simulation results are demonstrated to verify the optimality of the proposed IA under both ML decoding and MMSE decoding. 
	}

	\section{Problem Formulation}
	
	In this section, we give the problem formulation for the index assignment under ML decoding. Consider an $M$-level maximum entropy scalar quantizer
	characterized by a set of quantization levels (i.e.~the codebook)
	$\mathcal{Q}=\{q_0,\cdots, q_{M-1}\}$, where $q_i \in \mathbb{R}$ with $0\leq q_0<q_1<\ldots<q_{M-1}$. The maximum entropy quantizer outputs each quantization level with an equal probability of $1/M$. An $M$-PSK constellation is defined as
	\begin{equation}\label{constellation_psk}
		{\mathcal
			S}\triangleq \{s_k=e^{j2\pi k/M}| k=0,1,\ldots,M-1\}\,,
	\end{equation}
	where $j\triangleq \sqrt{-1}$. To describe the nonbinary index assignment for $M$-PSK, let us define a vector 
	\begin{equation}\label{mapping} 
		\bm{\pi}=[\pi_0,\pi_1,\ldots,\pi_{M-2},\pi_{M-1}]\,.
	\end{equation}
	It is a permutation of the quantization indices, i.e., $\pi_k \in \{0,\cdots,M\!-\!1\}$. The IA is a bijective mapping between the quantizer and the constellation in a way that each quantization level $q_{\pi_k}$ is assigned to a distinct $M$-PSK symbol $s_k$.
	
	Each quantization level is modulated as an $M$-PSK symbol following the bijective mapping $\bm{\pi}$, and the $M$-PSK symbol is then transmitted over a memoyless channel with an additive noise following a symmetrically decreasing probability density function (such as the AWGN channel). At the receiver, an $M$-PSK demodulator detects the most likely transmitted index based on the received signal and the quantizer decoder reconstructs the source symbol by producing the quantization level corresponding to the detected index. The channel MSD is defined as
	\begin{equation}\label{nmsd}
		D_C(\bm{\pi})=\frac{1}{M}\sum_{i=0}^{M-1}\sum_{j=0}^{M-1}P(s_j|s_i)(q_{\pi_i}-q_{\pi_j})^2\,,
	\end{equation}
	where $P(s_j|s_i)$ is the probability that $M$-PSK symbol $s_j$ is detected conditioned on $s_i$ is transmitted. The optimal IA problem is to find $\bm \pi$ that minimizes $D_C(\bm{\pi})$ over the set of all possible permutations. It is worth noting that the zigzag mapping \cite{chan2005index} is produced by the permutation 
	\begin{equation*}
		\bm \pi_{zz} \triangleq [0,1,3,\ldots,M-1,\ldots,6,4,2]\,.
	\end{equation*}
	
	Note that transition probabilities of the channel satisfy
	\begin{equation}\label{sym_chan}
		\sum_{i=0}^{M-1}P(s_j|s_i)=\sum_{j=0}^{M-1}P(s_j|s_i)=1\,.
	\end{equation}
	The channel MSD can be written as
	\begin{equation}\label{cmsd}
		D_C(\bm{\pi})=\frac{2}{M}\sum_{i=0}^{M-1}q_i^2-\frac{2}{M}\sum_{i=0}^{M-1}\sum_{j=0}^{M-1}q_{\pi_i}q_{\pi_j}P(s_j|s_i)\,.
	\end{equation}
	For a given quantizer, the first term of (\ref{cmsd}) is fixed. By defining $p_{i,j}\triangleq P(s_j|s_i)$, the nonbinary index assignment problem can be formulated as 
	\begin{equation}\label{opt_pi}
		\max_{\bm{\pi}\in {\mathcal{P}}}\sum_{i=0}^{M-1}\sum_{j=0}^{M-1}q_{\pi_i}q_{\pi_j}p_{i,j}\,,
	\end{equation}
	where $\mathcal{P}$ denotes the set of all possible $M$-ary permutations.

	\section{Optimal Index Assignment based on a Discrete Convolution-Rearrangement Inequality}

	For $M$-PSK transmission, the channel matrix $P$ of the resulted $M$-ary DMC always satisfies the following conditions
	\begin{subequations}\label{req_psk}
		\begin{numcases}{}
			p_{i,j}\geq 0,\label{req_psk_1}\\
			p_{i,j}= p_{i',j'},\quad d(i,j)=d(i',j'), \label{req_psk_2}\\
			p_{i,j}\geq p_{i',j'},\quad d(i,j)<d(i',j')\label{req_psk_3}\,,
		\end{numcases}
	\end{subequations}
	where $0\leq i,j,i'\!,\!j'\leq M\!-\!1$, and $d(i,j)$ is defined by
	\textcolor{black}{
		\begin{equation}\label{def_distance}
			d(i,j)\triangleq\min\{(i-j)\bmod M,M-\big((i-j)\bmod M\big)\}\,,
		\end{equation}
		where the $\bmod$ operation returns an integer between $0$ and $M-1$.
	}
	
	Note that $P$ is a non-negative circulant matrix, i.e., $p_{i,j}$ is a non-negative function involving two integers $i,j$ and its value depends on $d(i,j)$ only. For convenience of notation we define a monotonically decreasing function $k(d(i,j))\triangleq p_{i,j}$. Let us define a vector $\bm{x}$ by
	\begin{equation}
		\bm{x}\triangleq[x_{-m},x_{-m+1},\ldots,x_0,\ldots,x_{n-1},x_{n}]\,,
	\end{equation}
	where $m\triangleq\lfloor(M\!-\!1)/2\rfloor$, $n\triangleq\lfloor M/2\rfloor$ and $x_i\triangleq q_{\pi_{i+m}}$. Then (\ref{opt_pi}) can be rewritten in a discrete convolution form 
	
	\begin{equation}\label{opt_pi_zm}
		\max_{\bm{x}\in\mathcal{P_Q}}\sum_{i=-m}^{n}\sum_{j=-m}^{n}x_ik(d(i,j))x_j\,,
	\end{equation}
	where $\mathcal{P_Q}$ is the set of all orderings of quantizer codebook $\mathcal{Q}$.

	To solve the problem (\ref{opt_pi_zm}), we give a discrete convolution-rearrangement inequality in the following theorem. 
	\medskip
	\begin{Theorem}\label{inequality}
		Suppose $M$ is a positive integer. Let $\mathbb{Z}_M=\{\lceil\frac{1-M}{2}\rceil,\lceil\frac{1-M}{2}\rceil+1,\ldots,\lfloor\frac{M}{2}\rfloor\}$. Let $k$ be a decreasing non-negative function on $[0,\infty)$ and let $d(i,j)$ be defined by (\ref{def_distance}). For any two non-negative functions $f$ and $g$ on $\mathbb{Z}_M$, we have
		\begin{equation}\label{rear_ineq}
			\begin{aligned}
				\sum_{i,j\in\mathbb{Z}_M}\!\!f(i)k(d(i,j))g(j)\leq \sum_{i,j\in\mathbb{Z}_M}\!\!f^{\star}(i)k(d(i,j))g^{\star}(j)\,,
			\end{aligned}
		\end{equation}
		where $f^{\star}$ is a discrete symmetric decreasing rearrangement of $f$ such that $f^{\star}$ is derived from a permutation on the set of function values of $f$ and satisfies
		\begin{small}
			\begin{equation*}
				\left\{
				\begin{aligned}
					&f^{\star}(0)\geq f^{\star}(1)\geq f^{\star}(-1)\geq f^{\star}(2)\geq f^{\star}(-2)\geq\ldots\\
					&\;\;\;\;\;\;\;\;\;\;\;\;\;\;\;\;\;\;\;\;\;\;\;\;
					\geq f^{\star}(\lfloor\frac{M}{2}\rfloor)\geq f^{\star}(\lceil\frac{1-M}{2}\rceil),\quad M\text{ is odd}\,,\\
					&f^{\star}(0)\geq f^{\star}(1)\geq f^{\star}(-1)\geq f^{\star}(2)\geq f^{\star}(-2)\geq\ldots\\
					&\;\;\;\;\;\;\;\;\;\;\;\;\;\;\;\;\;\;\;\;\;\;\;\;
					\geq f^{\star}(\lceil\frac{1-M}{2}\rceil) \geq f^{\star}(\lfloor\frac{M}{2}\rfloor),\quad M\text{ is even}\,.
				\end{aligned}
				\right.
			\end{equation*}
		\end{small}
	\end{Theorem}
	
	\medskip
	For ease of understanding, we provide a simple example with $M=6$. Let us consider a non-negative monotonically decreasing function $k(d(i,j))=e^{-{d(i,j)}}$ with integers $i,j\in[-2,3]$ and consider $f$ and $g$ as
	\begin{equation*}
		\left\{
		\begin{aligned}
			&f(i)=i^2,\\
			&g(i)= M+2i,
		\end{aligned}
		\right. \quad i=-2,-1,\ldots,2,3\,.
	\end{equation*}
	Denotes by $\bm f$ a vector whose $i$-th element is $f(i)$,  in which $i=-2,-1,\ldots,2,3$. Two vectors associated with $f$ and $g$ are
	\begin{equation*}
		\begin{aligned}
			\bm f=[4,1,0,1,4,9],\quad \bm g=[2,4,6,8,10,12]\,.
		\end{aligned}
	\end{equation*}
	We want to find orderings (i.e., permutations) of $\bm f$ and $\bm g$ that maximize the discrete convolution on the left-hand side of (\ref{rear_ineq}). Note that there are $M$ elements in $\bm f$ and $\bm g$, and there are $M!$ orderings of $\bm f$ and $\bm g$, respectively. Among all of the $M!\times M!$ orderings, the one following
	\begin{equation*}
		\begin{aligned}
			\bm f^{\star}=[1,4,9,4,1,0],\quad \bm g^{\star}=[4,8,12,10,6,2]\,,
		\end{aligned}
	\end{equation*}
	gives the maximum discrete convolution value. 

	\begin{Remark}
		\textcolor{black}{
			An anonymous reviewer points out that Theorem~\ref{inequality} we derived is a rediscovery of \cite[Theorem 5.1]{pruss1998discrete}. Due to space limitation, readers can find our proof in the full version of this paper \cite{preprint_proof}.
		}
	\end{Remark}
	
	The major difference of Theorem~\ref{inequality} and the Hardy-Littlewood convolution-rearrangement inequality \cite[Theorem 371]{hardy1952inequalities} is the definition of $d(i,j)$. For Theorem~\ref{inequality}, if we consider a graph composed of $M$ points joined together in a circle, then $d(i,j)$ agrees with the graphic distance between the $i$-th and the $j$-th vertices on the circle. Therefore, $d(i,j)$ is a periodic function of $i-j$ with period $M$. However, in \cite[Theorem 371]{hardy1952inequalities} $d(i,j)$ is symmetrically increasing with $i-j$. It is worth noting that the Hardy-Littlewood convolution-rearrangement inequality \cite[Theorem 371]{hardy1952inequalities} is a special case of Theorem~1 and can be obtained by letting the number of points on the discrete circle graph tend to infinity.
	
	According to Theorem~\ref{inequality}, the objective function of (\ref{opt_pi_zm}) achieves its maximum when $\bm{x}$ is ordered as
	\begin{equation}
		x_{0}\geq x_{1}\geq x_{{-1}}\geq x_{2}\geq x_{{-2}}\geq\ldots\,,
	\end{equation}
	where $\ldots$ indicates that we keep on going until we exhaust all integers in $\mathbb{Z}_M$. 
	
	The optimal solution for (\ref{opt_pi}) can be consequently found as
	\begin{equation}
		\left\{
		\begin{aligned}
			&\bm{\pi}=[0,2,\ldots,{M-1},{M-2},\ldots,1],\quad \text{for odd } M,\\
			&\bm{\pi}=[1,3,\ldots,{M-1},{M-2},\ldots,0],\quad \text{for even } M.
		\end{aligned}
		\right.
	\end{equation}
	Finally, we have Theorem~\ref{IA} for the optimal IA for $M$-PSK under ML decoding.
	\medskip
	\begin{Theorem}\label{IA}
		For maximum entropy scalar quantizers and $M$-PSK transmission over a memoryless channel with additive noise following a symmetrically decreasing probability density function, the optimal IA under ML decoding for minimizing channel MSD is
		\begin{equation}\label{opt_ia}
			\left\{
			\begin{aligned}
				&[q_0,q_2,\ldots,q_{M-1},q_{M-2},\ldots,q_1],\quad \text{for odd } M,\\
				&[q_1,q_3,\ldots,q_{M-1},q_{M-2},\ldots,q_0],\quad \text{for even } M.
			\end{aligned}
			\right.
		\end{equation}
	\end{Theorem} 
	\medskip
	
	In \cite{chan2005index}, a set of distortion-preserving transforms for $M$-PSK are introduced. Given any IA, cyclically shifting the indices to the right, i.e., $[q_1,q_3,\ldots,q_{2},q_{0}]\rightarrow[q_0,q_1,q_3,\ldots,q_{2}]$, does not change the channel MSD. Besides, a reflection operation $[q_2,q_4,\ldots,q_{1},q_{0}]\rightarrow[q_0,q_1,\ldots,q_{4},q_{2}]$ does not influence the channel MSD. Note that the IA for even $M$ in (\ref{opt_ia}) can be transformed to the zigzag mapping \cite{chan2005index} by a cyclic shift operation, and the IA for odd $M$ in (\ref{opt_ia}) can be transformed to the zigzag mapping by a reflection operation and a cyclic shift operation. Therefore, the zigzag mapping under ML decoding is proved to be optimal for all SNRs. 
	
	\section{Optimal Index Assignment for $M$-PSK under MMSE Decoding}
	
	In this section, the optimal IA for $M$-PSK under MMSE decoding is investigated. Different from the ML decoder that maps the detected $M$-PSK symbol back to a quantization level following the IA, we can alternatively consider an MMSE decoder which computes and outputs $y_j$ based on detected symbol ${s}_j$ by
	\begin{equation}
		\begin{aligned}
			y_j=E(q|s_j)=\frac{\sum_{k=0}^{M-1}\frac{1}{M}q_{\pi_k}P(s_j|s_k)}{\sum_{k=0}^{M-1}\frac{1}{M}P(s_j|s_k)}\,,
		\end{aligned}
	\end{equation}
	and the channel MSD is
	\begin{equation}
		D_C(\bm\pi) =\frac{1}{M}\sum_{i=0}^{M-1}\sum_{j=0}^{M-1}P(s_j|s_i)(q_{\pi_i}-y_j)^2\,.
	\end{equation}
	According to (\ref{sym_chan}), the channel MSD is formulated as
	\begin{equation}
		\begin{aligned}
			D_C(\bm \pi)&=\frac{1}{M}\Big(\sum_{i=0}^{M-1}q_i^2-2\sum_{j=0}^{M-1}\sum_{i=0}^{M-1}P(s_j|s_i)q_{\pi_i}y_j+\sum_{j=0}^{M-1}y_j^2\Big)\\
			&=\frac{1}{M}\sum_{i=0}^{M-1}q_i^2-\frac{1}{M}\sum_{j=0}^{M-1}\big(\sum_{k=0}^{M-1}q_{\pi_k}P(s_j|s_k)\big)^2\,.
		\end{aligned}
	\end{equation}
	Letting $p_{k,j}= P(s_j|s_{k})$, then minimizing the channel MSD is equivalent to
	\begin{equation}\label{ia_mmse_1}
		\max_{\bm \pi\in \mathcal{P}}\sum_{j=0}^{M-1}\big(\sum_{k=0}^{M-1}q_{\pi_k}p_{k, j}\big)^2.
	\end{equation}
	And it can be simplified as
	\begin{equation}\label{pro_mmse}
		\max_{\bm \pi\in \mathcal{P}}\sum_{i=0}^{M-1}\sum_{j=0}^{M-1}q_{\pi_i}q_{\pi_j}h_{i,j}\,,
	\end{equation}
	where $H=PP^T$. 
	
	The similarity between (\ref{pro_mmse}) and (\ref{opt_pi}) gives us the insight to solve (\ref{pro_mmse}) by Theorem~\ref{inequality}. For this purpose, we investigate the property of $H$ and get the following lemma. 
	
	\medskip
	\begin{Lemma}
		Suppose that the conditions in (\ref{req_psk}) hold for two square matrices $Q$ and $R$. Then the conditions in (\ref{req_psk}) also hold for the matrix $QR^T$. 
	\end{Lemma}
	\medskip
	\begin{IEEEproof}
		The condition (\ref{req_psk_1}) holds for the matrix $QR^T$ since the product of two non-negative matrices is also non-negative.
		
		Let us define $H\triangleq QR^T$. Note that $Q$ and $R$ are symmetric circulant matrices since (\ref{req_psk_2}) holds for them. According to \cite{horn2012matrix}, $H$ is also a symmetric circulant matrix. This fact implies that (\ref{req_psk_2}) also holds for $H$. 
		
		To prove the condition (\ref{req_psk_3}), let us define two function compositions by $k_q(d(i,j))\triangleq q_{i,j}$ and $k_r(d(i,j))\triangleq r_{i,j}$ for $1\leq i,j\leq M$, where $k_r$ and $k_q$ are two monotonically decreasing functions, and $d(i,j)$ is defined by (\ref{def_distance}). Because of (\ref{req_psk_2}), the two function compositions can represent all entries in $Q$ and $R$. Entries of $H$ are computed by
		\begin{equation}\label{hij}
			h_{i,j}=\sum_{k=1}^{M}q_{i,k}r_{j,k}=\sum_{k=1}^{M}k_q(d(i,k))k_r(d(j,k))\,.
		\end{equation}
		The condition (\ref{req_psk_2}) holds iff the following condition
		\begin{equation}\label{suff_con}
			h_{i,j}\geq h_{i',j'},\quad \text{if } d(i',j')=d(i,j)+1
		\end{equation}
		holds. Defining $\Delta_{ii'jj'}\triangleq h_{i,j}-h_{i',j'}$. For the sake of convenience, subscripts of $\Delta$ will be omitted in the remainder of the paper. Then based on (\ref{hij}) we have
		\begin{equation}\label{proof_lemma}
			\begin{aligned}
				\Delta&=\sum_{k=1}^{M}k_q(d(i,k))k_r(d(j,k))-\sum_{k=1}^{M}k_q(d(i',k))k_r(d(j',k))\\
				&\overset{(a)}=\sum_{k=1}^{M}k_q(d(0,k))k_r(d(j-i,k))\\
				&\;\;\;\;-\sum_{k=1}^{M}k_q(d(0,k))k_r(d(j'-i',k))\,,
			\end{aligned}
		\end{equation}
		where equality $(a)$ follows from the fact that
		\begin{equation*}
			\begin{aligned}
				\sum_{k=1}^{M}k_q(d(i,k))k_r(d(j,k))&=\sum_{k=1-i}^{M-i}k_q(d(0,k))k_r(d(j\!-\!i,k))\,,
			\end{aligned}
		\end{equation*}
		\textcolor{black}{and $d(i,k)=d(i,k+M)$. }
		
		\textcolor{black}{
			According to (\ref{def_distance}), if $i$ and $j$ are integers between $1$ and $M$, $d(i,j)$ is equivalent to the following two cases
			\begin{equation*}
				d(i,j)=\left\{
				\begin{aligned}
					& |j-i|,&& |j-i| \leq M/2\,,\\
					& M-|j-i|,&& |j-i| > M/2\,.
				\end{aligned}
				\right.
			\end{equation*}
			Then the condition $d(i',j')=d(i,j)+1$ in (\ref{suff_con}) is equivalent to the following four cases
			\begin{small}
				\begin{equation}\label{two_cases}
					\left\{
					\begin{aligned}
						& |j'-i'|=|j-i|+1,&& |j-i| < M/2, |j'-i'| \leq M/2\,,\\
						& M-|j'-i'|=|j-i|+1,&& |j-i| < M/2, |j'-i'| \geq M/2\,,\\
						& |j'-i'|=M-|j-i|+1,&& |j-i| > M/2, |j'-i'| \leq M/2\,,\\
						& |j'-i'|=|j-i|-1,&& |j-i|> M/2, |j'-i'|\geq M/2\,.
					\end{aligned}
					\right.
				\end{equation}
			\end{small}For a given $|j-i|$, the first two cases have the same value of $d(i',j')$. Because of (\ref{req_psk_2}), the two cases also have the same value of $h_{i',j'}$. For proving (\ref{suff_con}), it is sufficient to examine one of them for $|j-i| < M/2$. Similarly, we just need to consider one of the last two cases for $|j-i|> M/2$.} 
		
		Then $\Delta$ can be sufficiently examined by the first and the last cases in (\ref{two_cases}). We give the proof of the first case. The last case can be proved similarly. Since $H$ is symmetric, let us assume $i-j\geq 0$ and $i'-j'\geq 0$ and get $j'-i'=j-i-1$. Then (\ref{proof_lemma}) is equivalent to
		\begin{equation}\label{proof_lemma2}
			\begin{aligned}
				\Delta&=\sum_{k=1}^{M}k_q(d(0,k))k_r(d(j-i,k))\\
				&\;\;\;-\sum_{k=1}^{M}k_q(d(0,k-1))k_r(d(j-i,k))\,,
			\end{aligned}
		\end{equation}
		which follows from the fact that
		\begin{equation*}
			\begin{aligned}
				&\sum_{k=1}^{M}k_q(d(0,k))k_r(d(j\!-\!i\!-\!1,k))\\
				&=\sum_{k=2}^{M+1}k_q(d(0,k-1))k_r(d(j-i,k))\,,
			\end{aligned}
		\end{equation*}
		and $d(i,k)=d(i,k+M)$. 
		
		Assume $M$ is even. Note that the case of odd $M$ can be proved similarly. From (\ref{proof_lemma2}) we have
		\begin{equation*}
			\begin{aligned}
				\Delta&=\sum_{k=1}^{M}\Big(k_q(d(0,k))-k_q(d(0,k\!-\!1))\Big)k_r(d(j\!-\!i,k))\\
				&=\sum_{k=1}^{\frac{M}{2}}\Big(k_q(d(0,k))-k_q(d(0,k\!-\!1))\Big)k_r(d(j\!-\!i,k))\\
				&\;\; +\sum_{k=\frac{M}{2}+1}^{M}\Big(k_q(d(0,k))-k_q(d(0,k\!-\!1))\Big)k_r(d(j\!-\!i,k))\\
				&\overset{(b)}=\sum_{k=1}^{\frac{M}{2}}\Big(k_q(d(0,k))-k_q(d(0,k\!-\!1))\Big)k_r(d(j\!-\!i,k))\\
				&\;\; +\sum_{k'=1}^{\frac{M}{2}}\Big(k_q(d(0,k'\!-\!1))-k_q(d(0,k'))\Big)k_r(d(j\!-\!i,1\!-\!k'))\\
				&=\sum_{k=1}^{\frac{M}{2}}\!\big(k_q(d(0,k))-k_q(d(0,k-1))\big)\\
				&\;\;\;\;\;\;\times \big(k_r(d(j-i,k))-k_r(d(j-i,1-k))\big)\,.
			\end{aligned}
		\end{equation*}
		where equality $(b)$ is obtained by letting $k'= M-k+1$.  It is obvious that 
		\begin{equation*}
			k_q(d(0,k))-k_q(d(0,k-1))\leq 0,\quad 1\leq k\leq M/2\,.
		\end{equation*}
		Beside, we have
		\begin{equation*}
			k_r(d(j-i,k))-k_r(d(j-i,1-k))\leq 0,\quad 1\leq k\leq M/2\,,
		\end{equation*}
		which follows from the fact that 
		\begin{equation*}
			d(j-i,k)\geq d(j-i,1-k),\quad 0\leq i-j<M/2\,.
		\end{equation*}
		Therefore, we have $\Delta\geq 0$. The condition (\ref{req_psk_3}) is proved consequently. 
	\end{IEEEproof}
	
	According to Lemma~1, the optimal IA problem for $M$-PSK under MMSE decoding can also be solved by the Theorem~\ref{inequality}. We have Theorem~3 for the optimal IA.
	\medskip
	\textcolor{black}{
		\begin{Theorem}\label{IA2}
			For maximum entropy scalar quantizers and $M$-PSK transmission over a memoryless channel with additive noise following a symmetrically decreasing probability density function, the zigzag mapping is the optimal IA under MMSE decoding.
		\end{Theorem}
	}
	\section{Numerical Results}

	To verify the optimality of the zigzag mapping, numerical results are demonstrated to compare the MSD performance of the zigzag mapping with the optimal mapping by the exhaustive search. Real-valued source symbols following a uniform distribution over $[0,1]$ are generated. The source symbols are then quantized by an $M$-level uniform scalar
	quantizer. The $M$-level quantized symbols are mapped to $M$-PSK symbols
	following an index assignment. $M$-PSK symbols are transmitted over the AWGN channel. After ML or MMSE decoding, the source data are reconstructed. To show that the zigzag mapping is optimal at all SNRs, we perform an exhaustive search for each simulated SNR separately.  
	
	To show the gain of the nonbinary index assignment, we also compare its performance with the state-of-the-art binary counterpart. To modulate the bits as $M$-PSK symbols, Gray code is used to minimize the bit error rate of the resultant BSC. In binary index assignment design for maximum entropy scalar quantizers and the BSC, the NBC is known to be optimal under both ML decoding \cite{mclaughlin1995optimal} and MMSE decoding \cite{farber2004quantizers}. Therefore, the NBC-Gray mapping is considered as the binary counterpart. To make the equivalent BSC from the $M$-PSK transmission memoryless, we assume an ideal bit interleaver to eliminate the correlation among all Gray coded bits.
	
\begin{figure}[!b]
		\centering
		\includegraphics[width=2.9in]{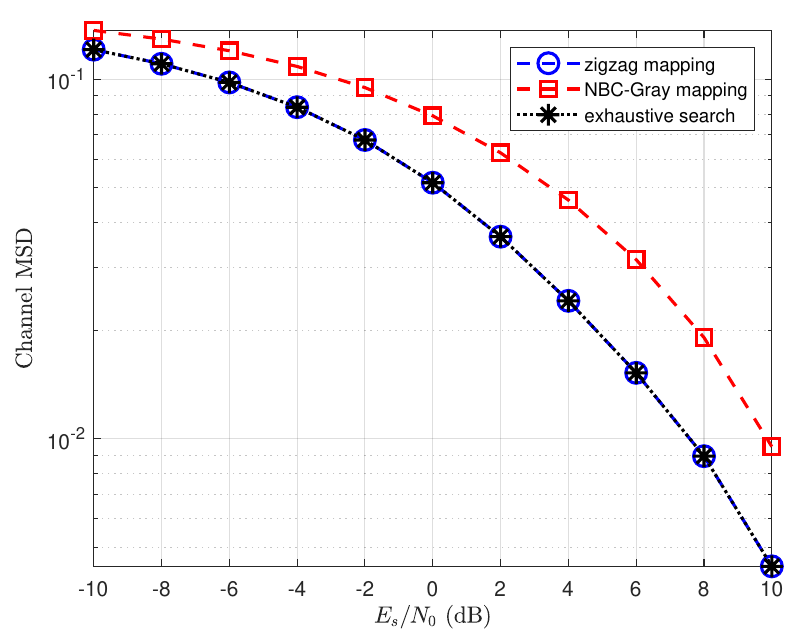}
		\caption{Channel MSD of the zigzag mapping, the NBC-Gray mapping and the optimal mapping exhaustively search for each SNR for $8$-PSK (i.e., $M=8$) under ML decoding in AWGN channel.} \label{8PSK_unis_uniq_ML}
	\end{figure}

\begin{figure}[!t]
		\centering
		\includegraphics[width=2.9in]{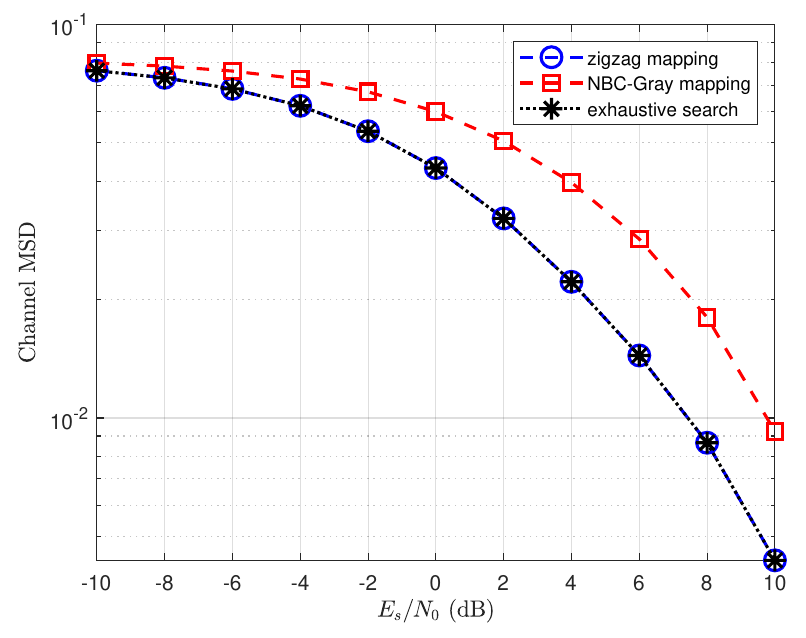}
		\caption{Channel MSD of the zigzag mapping, the NBC-Gray mapping and the optimal mapping exhaustively search for each SNR for $8$-PSK (i.e., $M=8$) under MMSE decoding in AWGN channel.}\label{8PSK_unis_uniq_MMSE}
	\end{figure}

	Figure~\ref{8PSK_unis_uniq_ML} and Figure~\ref{8PSK_unis_uniq_MMSE} show channel MSD performances under ML decoding and MMSE decoding, respectively. To verify the optimality of the zigzag mapping at all SNRs, we also provide results at low SNRs. Here we only provide the results for $M=8$ since the search space (which consists of $M!$ candidates) has a size that grows exponentially with $M$.
	
	From the figures, the performances of the zigzag mapping and the exhaustive search mapping are always the same. We also checked exhaustive search IAs for all simulated SNRs. They are the same as the zigzag mapping or can be transformed to the zigzag mapping by distortion-preserving transforms in \cite{chan2005index}. Besides, the performance gain of the optimal zigzag mapping over that of the NBC-Gray mapping is significant. The SNR gains are up to around $3$ dB under both ML and MMSE decoding. It is also worth noting that the channel MSD under MMSE decoding outperforms the one under ML decoding, especially at low SNRs.

	\section{Conclusion}

	A discrete convolution-rearrangement inequality was rediscovered. The inequality was applied to settle the optimal nonbinary index assignment problem for the $M$-level maximum entropy scalar quantizer and the $M$-PSK over a memoryless channel. For both ML and MMSE decoding, the zigzag mapping has been proved to be optimal for all SNRs. Simulation results were provided to verify the optimality of the zigzag mapping and to show the gain over the conventional binary index assignment. Further research on the application of rearrangement inequalities to address the IA and other problems in coding theory is a promising direction.

	\appendix
	The discrete decreasing rearrangement $f^{\star}$ and $g^{\star}$ is equivalent to the fact that
	\begin{small}
		\begin{equation}\label{codition_max}
			\left\{
			\begin{aligned}
				&f^{\star}(r)-f^{\star}(r')\geq 0,\quad \text{if }  |r'|>|r|, \text{ or } \text{if } r'=-r<0\,,\\
				&g^{\star}(s)-g^{\star}(s')\geq 0,\quad \text{if } |s'|>|s|,\text{ or } \text{if } s'=-s<0\,.
			\end{aligned}
			\right.
		\end{equation}
	\end{small}To prove the ordering following (\ref{codition_max}) gives the maximum, we define an operation $\Omega_p(f,g)$ that swaps all pairs 
	\begin{small}
		\begin{equation}\label{pair_even}
			\big(f(p-i),f(p+i)\big),\ \big(g(p-j),g(p+j)\big),\quad i,j=1,2,3,\ldots,
		\end{equation}
	\end{small}or in pairs 
	\begin{small}
		\begin{equation}\label{pair_odd}
			\big(f(p\!-\!i),f(p+i+1)\big),\ \big(g(p-j),g(p+j+1)\big),\quad i,j=0,1,2,\ldots,
		\end{equation}
	\end{small}which do not satisfy conditions in (\ref{codition_max}). Note that $f(r),g(s)$ are set as $0$ when $r,s$ out of the scope of $[-m,n]$. It is worth pointing out that starting from any $f$ and $g$, the operation $\Omega_p(f,g)$ can be applied for different $p$ iteratively, until we get $f^{\star}$ and $g^{\star}$. Therefore, a sufficient condition of Theorem 1 is that $\Omega_p$ always introduces non-negative increment. 
	
	To prove Theorem 1, we give the following lemma as the sufficient condition of Theorem 1.
	\begin{Lemma}\label{lemma_theorem1}
		For any $f$ and $g$, the operation $\Omega_p(f,g)$ for arbitrary $p$ can always introduce non-negative increment to the discrete convolution on thee left hand side of (\ref{rear_ineq}).
	\end{Lemma}
	\begin{IEEEproof}
		Note that (\ref{pair_even}) and (\ref{pair_odd}) always involve different pairs. We can prove Lemma~\ref{lemma_theorem1} for them separately. Here we give the proof for pairs in (\ref{pair_odd}) and the proof for pairs in (\ref{pair_even}) can be done following a similar procedure. In the proof we also assume $M$ is an even number, the case for odd $M$ also can be done similarly.
		
		Given any $p$, then $i,j$ in (\ref{pair_odd}) satisfy that 
		\begin{small}
			\begin{equation}
				\left\{
				\begin{aligned}
					&|p-i|\leq |p+i+1|,\ |p-j|\leq |p+j+1|,\quad p\geq 0,\\
					&|p-i|\geq |p+i+1|,\ |p-j|\geq |p+j+1|,\quad p<0. 
				\end{aligned}
				\right.
			\end{equation}
		\end{small}For given $f$ and $g$, let us denote by $I_w$ and $J_w$ the set of $i$ and $j$ for which
		\begin{small}
			\begin{equation*}
				\left\{
				\begin{aligned}
					&f(p-i)<f(p+i+1), \ g(p-j)<g(p+j+1),\quad p\geq 0,\\
					& f(p-i)>f(p+i+1), \ g(p-j)>g(p+j+1),\quad p< 0. 
				\end{aligned}
				\right.
			\end{equation*}
		\end{small}Note that the pairs corresponding to $i\in I_w$ and $j\in J_w$ do not satisfy (\ref{codition_max}). We also define the set of $i,j$ who satisify (\ref{codition_max}) as $I_r$ and $J_r$. Note that the union of $I_w$ and $I_r$ is the set of all possible values that $i$ can be, let us define it by $I$. Similarly, the union of $J_w$ and $J_r$ can also be defined as $J$. 
		
		Let us define $d$ as the value of the discrete convolution, i.e.,
		\begin{small}
			\begin{equation}\label{rear_ineq_proof}
				\begin{aligned}
					\chi=\sum_{r,s\in\mathbb{Z}_M}\!\!f(r)k(d(r,s))g(r)\,.
				\end{aligned}
			\end{equation}
		\end{small}According to the aforementioned definition of $I$ and $J$, (\ref{rear_ineq_proof}) can be divided into four partial sums
		\begin{small}
			\begin{equation}\label{rear_ineq_proof2}
				\begin{aligned}
					d
					&=\sum_{i\in I}\sum_{j\in J}\!\!\Big(k(d(i,j))\big(f(p\!-\!i)g(p\!-\!j)\!+\!f(p\!+\!i\!+\!1)g(p\!+\!j\!+\!1)\big)\\
					&\;\;\;\;\;\;\;\;+k(d(i\!+\!1,\!-\!j))\big(f(p\!-\!i)g(p\!+\!j\!+\!1)\!+\!f(p\!+\!i\!+\!1)g(p\!-\!j)\big)\Big)\\
					&=\sum_{i\in I_r}\sum_{j\in J_r}\!\!\Big(k(d(i,j))\big(f(p\!-\!i)g(p\!-\!j)\!+\!f(p\!+\!i\!+\!1)g(p\!+\!j\!+\!1)\big)\\
					&\;\;\;\;\;\;\;\;+k(d(i\!+\!1,\!-\!j))\big(f(p\!-\!i)g(p\!+\!j\!+\!1)\!+\!f(p\!+\!i\!+\!1)g(p\!-\!j)\big)\Big)\\
					&+\sum_{i\in I_w}\sum_{j\in J_w}\!\!\Big(k(d(i,j))\big(f(p\!-\!i)g(p\!-\!j)\!+\!f(p\!+\!i\!+\!1)g(p\!+\!j\!+\!1)\big)\\
					&\;\;\;\;\;\;\;\;+k(d(i\!+\!1,\!-\!j))\big(f(p\!-\!i)g(p\!+\!j\!+\!1)\!+\!f(p\!+\!i\!+\!1)g(p\!-\!j)\big)\Big)\\
					&+\sum_{i\in I_w}\sum_{j\in J_r}\!\!\Big(k(d(i,j))\big(f(p\!-\!i)g(p\!-\!j)\!+\!f(p\!+\!i\!+\!1)g(p\!+\!j\!+\!1)\big)\\
					&\;\;\;\;\;\;\;\;+k(d(i\!+\!1,\!-\!j))\big(f(p\!-\!i)g(p\!+\!j\!+\!1)\!+\!f(p\!+\!i\!+\!1)g(p\!-\!j)\big)\Big)\\
					&=\sum_{i\in I_r}\sum_{j\in J_w}\!\!\Big(k(d(i,j))\big(f(p\!-\!i)g(p\!-\!j)\!+\!f(p\!+\!i\!+\!1)g(p\!+\!j\!+\!1)\big)\\
					&\;\;\;\;\;\;\;\;+k(d(i\!+\!1,\!-\!j))\big(f(p\!-\!i)g(p\!+\!j\!+\!1)\!+\!f(p\!+\!i\!+\!1)g(p\!-\!j)\big)\Big)\,.
				\end{aligned}
			\end{equation}
		\end{small}Let us define the four partial terms corresponding to the four convolution ranges as $\chi_1$, $\chi_2$, $\chi_3$, and $\chi_4$. 
		
		To show the increment of $\Omega_p$ on $\chi$, its effect on each of the four terms need to be examined separately.
		It is clear that $\Omega_p$ has no influence on $\chi_1$. It is also trivial to check that $\chi_2$ is not affected by $\Omega_p$.
		Therefore, the only two partial sums need to be considered are $\chi_3$ and $\chi_4$.

		Let us define $d_3$ to be the increment produced by $\Omega_p$ on $\chi_3$, i.e., 
		\begin{small}
			\begin{equation}\label{sum_incre}
				\begin{aligned}
					d_3\triangleq&\sum_{i\in I_w}\sum_{j\in J_r}\!\!\Big(k(d(i,j))\big(f(p\!-\!i)g(p\!+\!j\!+\!1)\!+\!f(p\!+\!i\!+\!1)g(p\!-\!j)\big)\Big)\\
					&\;\;\;\;+k(d(i\!+\!1,\!-\!j))\big(f(p\!-\!i)g(p\!-\!j)\!+\!f(p\!+\!i\!+\!1)g(p\!+\!j\!+\!1)\big)\\
					-&\sum_{i\in I_w}\sum_{j\in J_r}\!\!\Big(k(d(i,j))\big(f(p\!-\!i)g(p\!-\!j)\!+\!f(p\!+\!i\!+\!1)g(p\!+\!j\!+\!1)\big)\\
					&\;\;\;\;+k(d(i\!+\!1,\!-\!j))\big(f(p\!-\!i)g(p\!+\!j\!+\!1)\!+\!f(p\!+\!i\!+\!1)g(p\!-\!j)\big)\Big)\,.
				\end{aligned}
			\end{equation}
		\end{small}Now the task becomes proving that (\ref{sum_incre}) is non-negative.
		
		Let us simplify $d_3$ in (\ref{sum_incre}) as
		\begin{small}
			\begin{equation}\label{sum_incre2}
				\begin{aligned}
					d_3=&\sum_{i\in I_r}(f(p\!+\!i\!+\!1)-f(p\!-\!i))\\
					&\times \sum_{j\in J_w}(k(d(i,j))-k(d(-i,j\!+\!1)))(g(p\!-\!j)-g(p\!+\!j\!+\!1))\,.
				\end{aligned}
			\end{equation}
		\end{small}Recall that $f(r),g(s)$ are $0$ if $r,s$ outside of the scope of $[-m,n]$. If $j\in J_w$, both $p-j$ and $p+j+1$ should be in the range of $[-m,n]$. Therefore, the range of $I_w,J_w$ should be
		\begin{small}
			\begin{equation}\label{range_J}
				0\leq i,j\leq m-|p|,\quad \text{if }i\in I_w,j\in J_w.
			\end{equation}
		\end{small}\indent Similarly, for $i\in I_r$, at least one of $p-i$ and $p+i+1$ should be in the $[-m,n]$. The range of $I_w,J_w$ should be
		\begin{small}
			\begin{equation}\label{range_Jr}
				0\leq i,j\leq m+|p|,\quad \text{if }i\in I_r,j\in J_r.
			\end{equation}
		\end{small}Then $d_3$ in (\ref{sum_incre2}) can be further divided into two terms
		\begin{small}
			\begin{equation}\label{divide_sum}
				\begin{aligned}
					d_3=&\sum_{0\leq i\leq m-|p|,\atop i\in I_r}(f(p\!+\!i\!+\!1)-f(p\!-\!i))\\
					&\;\;\;\;\;\times\sum_{j\in J_w}(k(d(i,j))-k(d(-i,j\!+\!1)))(g(p\!-\!j)-g(p\!+\!j\!+\!1))\\
					&\;\;\;+\sum_{{{m-|p|+1\leq i\leq m+|p|}}}\!\!\!\!\!\!\!\!\!\!\!\!(f(p\!+\!i\!+\!1)-f(p\!-\!i))\\
					&\;\;\;\;\;\;\;\times\sum_{j\in J_w}(k(d(i,j))-k(d(-i,j\!+\!1)))(g(p\!-\!j)-g(p\!+\!j\!+\!1)).
				\end{aligned}
			\end{equation}
		\end{small}Note that the second term does not exist if $p=0$. And all $i$ in the range of the second term satisfy $i\in I_r$ because of (\ref{range_J}). 
		
		It is trivial that the first term in (\ref{divide_sum}) is non-negative.  When $p\neq 0$, the second term of (\ref{divide_sum}) also need to be considered, in which $m-|p|+1\leq i\leq m+|p|$. For each $m+1\leq i\leq m+|p|$, there is $m-|p|+1\leq (2m+1)-i\leq m$ such that
		\begin{small}
			\begin{equation}\label{pair_of_i}
				\left\{
				\begin{aligned}
					&d(i,j)=d(-((2m+1)-i),j+1),\\
					&d(-i,j+1)=d((2m+1)-i,j).\\
				\end{aligned}
				\right.
			\end{equation}
		\end{small}
		Therefore, the second term of (\ref{divide_sum}) can be written as
		\begin{small}
			\begin{equation}\label{second_term}
				\begin{aligned}
					&\sum_{\mathclap{\substack{m+1\leq i\leq m+|p|}}}\Big(\big(f(p\!+\!i\!+\!1)-f(p\!-\!i)\big)-\big(f(p\!+\!i'\!+\!1)-f(p\!-\!i')\big)\big)\\
					\times&\sum_{j\in J_w}\big(k(d(i,j))-k(d(-i,j+1))\big)\big(g(p\!-\!j)-g(p\!+\!j\!+\!1)\big)\,,
				\end{aligned}
			\end{equation}
		\end{small}where $i'= (2m+1)-i$ for notational convenience. 
		
		Note that $f(p-i)=f(p-i')=0$ when $p<0$ since both $p-i$ and $p-i'$ are outside the range $[-m,n]$. Similarly, $f(p+i+1)=f(p+i'+1)=0$ when $p>0$. Then for any $p\neq 0$ there is 
		\begin{small}
			\begin{equation*}
				\big(\big(f(p\!+\!i\!+\!1)-f(p\!-\!i)\big)-\big(f(p\!+\!i'\!+\!1)-f(p\!-\!i')\big)\big)\big)\big(g(p\!-\!j)-g(p\!+\!j\!+\!1)\big)\leq 0.
			\end{equation*}
		\end{small}We also can prove
		\begin{small}
			\begin{equation*}
				k(d(i,j))\leq k(d(-i,j+1)),\quad m+1\leq i\leq m+|p|,0\leq j\leq m-|p|
			\end{equation*}
		\end{small}by trivially check that
		\begin{small}
			\begin{equation*}
				d(i,j)\geq d(-i,j+1),\quad m+1\leq i\leq m+|p|,0\leq j\leq m-|p|\,.
			\end{equation*}
		\end{small}Therefore, (\ref{second_term}) is non-negative for an arbitrary $p$. Consequently, (\ref{divide_sum}) is proved to be non-negative for any $p$, i.e., $d_3$ is always non-negative. Similarly, the increment produced by $\Omega_p$ on $\chi_4$ can be proved to be non-negative. Finally, the increment introduced by $\Omega_p$ to all of $\chi_1,\chi_2,\chi_3,\chi_4$ are non-negative. The increment produced by $\Omega_p$ on $\chi$ for any $p$ is always non-negative. 
	\end{IEEEproof}
	
	\bibliographystyle{IEEEtran}
	\bibliography{zig}
	
\end{document}